\documentclass[12pt,preprint]{emulateapj}

\shorttitle{The Afterglow of XRF\,050416A}
\newcommand{\NHunits}{\mbox{$10^{21}~{\rm cm}^{-2}$}}
\newcommand{\flux}{erg cm$^{-2}$ s$^{-1}$}

\def\ltsima{$\; \buildrel < \over \sim \;$}
\def\simlt{\lower.5ex\hbox{\ltsima}}
\def\gtsima{$\; \buildrel > \over \sim \;$}
\def\simgt{\lower.5ex\hbox{\gtsima}}

\newcommand \beq {\begin{equation}}
\newcommand \eeq {\end{equation}}

\def\la{\hbox{\hspace{1.5mm}}\raise2pt
       \vbox{\hbox{$<$}}\lower2pt
       \vbox{\moveleft9.0pt\hbox{$\sim$ }}\hbox{\hskip 0.05mm}}
\def\ga{\hbox{\hspace{1.5mm}}\raise2pt
       \vbox{\hbox{$>$}}\lower2pt
       \vbox{\moveleft9.0pt\hbox{$\sim$ }}\hbox{\hskip 0.05mm}}
\def\npar{\hbox{\hspace{1.0mm}}
       \vbox{\hbox{$\parallel$}}
       \vbox{\moveleft6.3pt\hbox{$\not$}}\hbox{\hskip 0.5mm}}

\begin{document}

\title{SWIFT XRT OBSERVATIONS OF THE AFTERGLOW OF XRF~050416A}

\author{
VANESSA~MANGANO\altaffilmark{1},
VALENTINA~LA~PAROLA\altaffilmark{1},
GIANCARLO~CUSUMANO\altaffilmark{1},
TERESA~MINEO\altaffilmark{1},
DANIELE~MALESANI\altaffilmark{2},
JAROSLAW~DYKS\altaffilmark{3},
SERGIO~CAMPANA\altaffilmark{4},
MILVIA~CAPALBI\altaffilmark{5},
GUIDO~CHINCARINI\altaffilmark{4,6},
PAOLO~GIOMMI\altaffilmark{5}, 
ALBERTO~MORETTI\altaffilmark{4},
MATTEO~PERRI\altaffilmark{5},
PATRIZIA~ROMANO\altaffilmark{4},
GIANPIERO~TAGLIAFERRI\altaffilmark{4},
DAVID~N.~BURROWS\altaffilmark{7},
{\rm NEIL~GEHRELS}\altaffilmark{8},
OLIVIER~GODET\altaffilmark{9},
STEPHEN~T.~HOLLAND,\altaffilmark{8,10},
JAMIE~A.~KENNEA\altaffilmark{7}, 
KIM~L.~PAGE\altaffilmark{9},
JUDITH~L.~RACUSIN\altaffilmark{7},
PETER~W.A.~ROMING\altaffilmark{7},
BING~ZHANG\altaffilmark{11}}

\altaffiltext{1}{INAF -- Istituto di Astrofisica Spaziale 
                 e Fisica Cosmica Sezione di Palermo, 
                 via Ugo La Malfa 153, I-90146 Palermo, Italy; 
                 \email{vanessa@ifc.inaf.it}}
\altaffiltext{2}{International School for Advanced Studies (SISSA-ISAS), 
                 via Beirut 2-4, I-34014 Trieste, Italy}
\altaffiltext{3}{Centrum Astronomiczne im.~M.~Kopernika PAN, Toru{\'n}, Poland}
\altaffiltext{4}{INAF -- Osservatorio Astronomico di Brera, 
                 via Emilio Bianchi 46, I-23807 Merate (LC), Italy}
\altaffiltext{5}{ASI Science Data Center, Via Galileo Galilei, 
                 I-00044 Frascati (Roma), Italy}
\altaffiltext{6}{Universit\`a degli studi di Milano-Bicocca,
                 Dipartimento di Fisica, piazza delle Scienze 3, 
                 I-20126 Milano, Italy}
\altaffiltext{7}{Department of Astronomy \& Astrophysics, 525 Davey Lab., 
                 Pennsylvania State University, University Park, PA 16802, USA}
\altaffiltext{8}{NASA/Goddard Space Flight Center, Greenbelt, MD 20771, USA}
\altaffiltext{9}{Department of Physics and Astronomy, University of Leicester, 
                 Leicester LE1 7RH, UK}
\altaffiltext{10}{Universities Space Research Association, 10227 Wincopin Circle, Suite 221, Columbia, MD 21044 USA}
\altaffiltext{11}{Physics Department, University of Nevada Las Vegas, NV, USA}

\begin{abstract}

Swift discovered XRF~050416A with the Burst Alert Telescope 
and began observing it with its narrow field instruments only 64.5~s 
after the burst onset. 
Its very soft spectrum classifies this event as an X-Ray Flash.
The afterglow X-ray emission was monitored up to 74 days after 
the burst. 
The X-ray light curve initially decays very fast (decay slope
$\alpha\sim 2.4$), subsequently flattens ($\alpha\sim 0.44$)
and eventually steepens again ($\alpha\sim 0.88$), similar
to many X-ray afterglows.
The first and second phases end $\sim 172$ and $\sim 1450$~s 
after the burst onset, respectively.
We find evidence of spectral evolution from a softer emission 
with photon index $\Gamma \sim 3.0$ during the initial steep
decay, to a harder emission with $\Gamma \sim 2.0$ during
the following evolutionary phases. 
The spectra show intrinsic absorption in the host
galaxy with column density of $\sim6.8\times10^{21}$~cm$^{-2}$. 
The consistency of the initial photon index with the
high energy BAT photon index suggests that the initial 
fast decaying phase of the X-ray light curve may be 
the low-energy tail of the prompt emission. 
The lack of jet break signatures in the X-ray afterglow
light curve is not consistent 
with empirical relations between the source rest-frame peak 
energy and the collimation-corrected energy of the burst.
The standard uniform jet model can give a possible 
description of the XRF\,050416A X-ray afterglow for
an opening angle larger than a few tens {\rm of} degrees, 
although numerical simulations show that the late time 
decay is slightly flatter than 
expected from on-axis viewing of a uniform jet.
A structured Gaussian-type jet model with uniform Lorentz factor
distribution and viewing angle outside the Gaussian core is
another possibility{\rm,} although a full agreement with data 
is not achieved with the numerical models explored.

\end{abstract}


\section{INTRODUCTION}
\label{section:introduction}

The Swift Gamma-Ray Burst Explorer \citep{gehr}, successfully launched 
on 2004 November 20, is dedicated to the discovery and study 
of gamma-ray bursts (GRBs) and their X-ray and optical afterglows.
Its fast-pointing capability, compared with previous satellites,
allows us to repoint towards GRB sources approximately 100 s after the burst
detection by the Burst Alert Telescope (BAT; \citealp{barth05a}) 
and to study, for the first time, the early phases of the afterglow evolution.
Moreover, the very broad energy coverage allows a simultaneous study
of the phenomenon in the optical, soft and hard X-ray bands.

{\rm One of the main Swift results has been the direct observation
of the transition between the prompt and the afterglow emission.
A growing number of early rapidly-fading X-ray afterglows have been
observed by Swift, that have been successfully interpreted 
as the tail of the prompt GRB emission (e.g.
\citealp{taglia05,cusumano05b,barth05b,nousek05,chinca05,zhang05}).
The emergence of the true X-ray afterglow component has
been identified with a break in the X-ray light curve to
a less steep (often flat) decay rate, sometimes accompanied 
by spectral variation across the break. Further breaks in the 
X-ray light curve have been related to standard afterglow 
evolution. 
The light curve of the X-ray counterpart of XRF~050416A show
all these characteristic features.}

BAT detected and located 
{\rm XRF~050416A} on 2005 April 16, 11:04:44.5 UT,
at the coordinates RA$_{\rm J2000} = 12 ^{\rm h} 33 ^{\rm m} 57\fs6$, 
Dec$_{\rm J2000} = +21\degr 03\arcmin 10\farcs8$,
with an uncertainty of 3$\arcmin$ \citep{saka1,saka2}. 
The light curve showed a single peak followed by a small bump with duration
$T_{90}$=2.4$\pm0.2$~s, with most of the energy emitted in the $15-50$~keV band.
The time-averaged energy distribution was well described by a power law 
($N(E) \propto E^{-\Gamma}$) with photon index $\Gamma=3.1\pm0.2$ (90\% confidence level) 
\citep{saka4}.
The soft spectrum and the fact that the fluence {\rm in the $15-30$~keV energy band 
is $6.1\times 10^{-7}$ erg cm$^{-2}$, larger than the fluence in the 
$30-400$~keV band ($1.3\times 10^{-7}$ erg cm$^{-2}$)
classify} this event as an X-ray flash (XRF; \citealp{heise01,lamb}). 
A {\rm complete} ground analysis of the BAT data is presented in \citet{saka4}.
These authors found that the best fit for the average energy distribution 
of the burst over the $T_{90}$ interval is given by a Band model \citep{band},
with peak energy $E_{\rm p} = 15.6^{+2.3}_{-2.7}$~keV, low-energy spectral
slope $\alpha_{\rm Band} = -1$ (fixed) and high-energy slope
$\beta_{\rm Band} < -3.4$ (68\% confidence level). This represents a 3.1~$\sigma$ 
improvement with respect to a simple power law fit 
{\rm ($\Gamma = 3.1 \pm 0.2$ over the 14$-$150 keV energy band)}. 
\citet{saka4} also showed that spectral hard-to-soft evolution was present 
during the BAT observation, with the spectrum becoming considerably 
softer at the end of each peak,
{\rm and estimated  an isotropic energy $E_{\rm iso} \sim 1.2 \times 10^{51}$~erg.}

Following the burst detection,
the satellite executed an immediate slew and promptly began collecting data
at 11:05:49 UT (64.5~s after the trigger) 
with the Ultraviolet/Optical Telescope  (UVOT; \citealp{rom05})
and at 11:06:00.6 UT (i.e. 76.1~s after the trigger) with the X-Ray Telescope 
(XRT; \citealp{bur05a}). 

In the first 100~s of observation UVOT revealed a new source in the $V$ 
filter at RA$_{\rm J2000} = 12^{\rm h} 33^{\rm m} 54\fs56$, 
Dec$_{\rm J2000} = +21\degr 03\arcmin 27\farcs3$ 
{\rm (with an uncertainty radius of $0\farcs56$ \citep{holland06})},
with magnitude $V=19.38$ \citep{scha1}. Starting with data taken 207~s after
the trigger, the source was also detected in the $U$ and $B$ bands, with magnitudes
$U=19.34\pm0.20$ and $B=19.85\pm0.20$~mag. There was no further detection in the $V$
band down to the 5~$\sigma$ limiting magnitude of 19.57~mag \citep{scha2}, 
173~s after the first $V$-band image. A fading source was also detected at 193~nm 
(UVW2 filter), placing an upper limit of 1 to the GRB redshift \citep{fox}.
The results and implications of the UVOT observations are discussed 
in {\rm \cite{holland06}}. 

Ground-based optical, NIR, and radio follow-up observations were performed with
several instruments. 
A fading source was detected with the ANU 2.3~m telescope in the $R$ band 
\citep{ander}, with the Palomar 200-inch Hale Telescope 
in the $K_{\rm s}$ band $\sim 6$~min after the BAT trigger \citep{cenko2}, 
with the MAO telescope in the $R$ band ($R = 20.85\pm0.12$, with a
900~s exposure, 11~hr after the trigger; \citealp{kahh}), and, marginally, 
with the KAIT in a 60~s $I$-band image, 7.4~min after the trigger \citep{li}. 
A late observation performed with the MAGNUM telescope equipped with the
MIP dual-beam optical-NIR imager detected the afterglow with $R = 21.3$~mag, 
12.2~hr after the trigger.
Spectra of the host galaxy of XRF~050416A were taken with the 
Low Resolution Imaging Spectrometer mounted 
on the 10-m Keck I telescope \citep{cenko3}. 
The spectrum indicates that the host galaxy is faint and blue with a large
amount of ongoing star formation. Spectral analysis 
revealed several emission lines including [OII], H$\beta$, H$\gamma$,
and H$\delta$, at a redshift $z=0.6535 \pm 0.0002$.
This is consistent with the prediction of 
\citet{fox} based on the afterglow detection in
the Swift UVOT UVW2 filter. XRF~050416A is thus one of the
closest {\rm long} GRBs discovered by Swift.

At radio frequencies, no source was detected with the VLA down to a limiting flux of
260~$\mu$Jy at 8.46~GHz 37~min after the burst \citep{frail} or with the Giant
Meter-wave Radio Telescope at 1280~MHz $\sim 9$~d after the burst (placing an
upper limit of 94~$\mu$Jy; \citealp{ish}), while a source with flux density
$260\pm55~\mu$Jy was detected with the VLA at 4.86~GHz, 5.6~d after the
burst \citep{sode}.

In the following, we report on the analysis of the prompt emission 
and of the X-ray afterglow observed by Swift. 
Details on the follow-up XRT observations and the XRT data reduction 
are described in \S~\ref{xrtdata}; 
the temporal and spectral analysis results are reported in \S~\ref{xrtanalysis}. 
In \S~\ref{discussion} we discuss our results.
Conclusions are drawn in \S~\ref{section::conclusions}. 
{\rm Finally, in the Appendix we will show how
our interpretation of the early XRT light curve as the tail of
the prompt emission could be reconciled with the report
of a peak energy $E_{\rm p} = 15.6^{+2.3}_{-2.7}$~keV 
by \citet{saka4} and the observed hardness evolution
of the BAT light curve.}

Throughout this paper the quoted uncertainties are given at the 90\%
confidence level for one interesting parameter, unless otherwise specified.
We also adopt the notation $F(\nu,t) \propto t^{-\alpha} \nu^{-\beta}$
for the afterglow monochromatic flux as a function of time, with $\nu$
representing the frequency of the observed radiation and with the energy
index $\beta$ related to the photon index $\Gamma$ according to
$\beta = \Gamma-1$. 

\begin{deluxetable*}{ccccccc}
\tablecolumns{6}
\tabletypesize{\scriptsize}
\tablecaption{XRT observation log of XRF~050416A}
\tablewidth{0pt}
\tablehead{
\colhead{\tiny Obs. \# \normalsize} &
\colhead{\tiny Sequence } &
\colhead{\tiny Mode \normalsize} &
\colhead{\tiny Start time \tiny(UT)\normalsize} &
\colhead{\tiny Start time \normalsize} &
\colhead{\tiny Exposure \normalsize}\\
\colhead{} &
\colhead{} &
\colhead{} &
\colhead{\tiny (yyyy-mm-dd hh:mm:ss.s)\normalsize } &
\colhead{\tiny(s since trigger)\normalsize} &
\colhead{\tiny(s)\normalsize} 
}
\startdata
  1& 00114753000 & LR\tablenotemark{a} & 2005-04-16 11:05:49.0 &  64.5 & 8.2   \\ 
  1& 00114753000 & IM & 2005-04-16 11:06:00.6 &      76.1 & 2.5   \\ 
  1& 00114753000 & WT & 2005-04-16 11:06:08.6 &      84.1 & 9.6   \\ 
  1& 00114753000 & PC & 2005-04-16 11:06:18.2 &      93.8 & 57360 \\ 
  2& 00114753001 & PC & 2005-04-18 14:27:56.9 &  184992.4 & 35084 \\
  3& 00114753003 & PC & 2005-04-26 00:53:32.7 &  827328.2 & 21279 \\
  4& 00114753004 & PC & 2005-04-28 01:07:03.8 & 1000939.3 & 20710 \\
  5& 00114753005 & PC & 2005-05-02 00:24:42.2 & 1343997.7 & 6892  \\
  6& 00114753006 & PC & 2005-05-03 00:33:40.8 & 1430936.3 & 5696  \\
  7& 00114753008 & PC & 2005-05-08 16:38:29.5 & 1920825.0 & 16702 \\
  8& 00114753009 & PC & 2005-05-13 01:08:16.3 & 2297011.8 & 22628 \\
  9& 00114753010 & PC & 2005-05-14 01:13:22.3 & 2383717.8 & 29625 \\
 10& 00114753011 & PC & 2005-05-25 04:03:25.2 & 3344320.7 & 23897 \\
 11& 00114753012 & PC & 2005-05-26 02:50:04.0 & 3426320.5 & 692   \\
 12& 00114753013 & PC & 2005-05-27 01:06:28.8 & 3506504.3 & 5077  \\
 13& 00114753014 & PC & 2005-05-29 01:18:21.3 & 3680016.8 & 19254 \\
 14& 00114753018 & PC & 2005-06-21 00:46:21.1 & 5665296.6 & 27979 \\
 15& 00114753019 & PC & 2005-06-22 00:46:54.2 & 5751729.7 & 24071 \\
 16& 00114753020 & PC & 2005-06-23 00:55:53.8 & 5838669.3 & 24073 \\
 17& 00114753021 & PC & 2005-06-25 01:07:54.8 & 6012190.3 & 14581 \\
 18& 00114753022 & PC & 2005-06-28 00:04:59.1 & 6267614.6 & 13215 \\
 19& 00114753023 & PC & 2005-06-29 00:05:00.2 & 6354015.7 & 11435 \\
\enddata
\label{log}
\tablenotetext{a}{The LR observation refers to the settling data set acquired during the slew
to the BAT coordinates. Only the time interval in which the X-ray countrpart of XRF~050416A is
clearly detected is considered.}
\end{deluxetable*}

\section{XRT OBSERVATIONS AND DATA REDUCTION}
\label{xrtdata}

The Swift XRT is designed to perform automated observations of newly discovered
bursts in the $0.2-10$~keV band.  
Four different read-out modes have been implemented, each dependent on the 
count rate of the observed sky region. The transition between two  modes is 
automatically performed  on board
(see \citealp{hill,hill05} for a detailed description of XRT observing modes). 

XRT was on target 76.1~s after the BAT trigger. 
It was operating in auto state and went through the standard sequence 
of observing modes, 
slewing to the GRB field of view in Low Rate Photodiode (LR) mode, 
taking a 2.5~s frame in Image (IM) mode followed by a LR frame (1.3~s), 
and 8 Windowed Timing (WT) mode 
frames (9.6~s), and then correctly switching to Photon Counting (PC) mode 
for the rest of the orbit. 
XRT was not able to automatically detect the source centroid on-board
because of its low intensity, but ground analysis revealed a fading
object identified as the X-ray counterpart of XRF~050416A \citep{cusumano05a}.
XRF~050416A was then observed {\rm intermittently over 29} consecutive 
orbits for a total exposure time of 57\,454~s. 
During the 8th orbit (starting $\sim35$~ks after the
trigger), the brightening of a column of flickering pixels caused uncontrolled
mode-switching between WT and PC modes. The WT data from this orbit are not 
usable because of their very low signal to noise (S/N) ratio.
XRF~050416A was further observed several times up to 74~d later 
in PC mode. The observation log is presented in Table \ref{log}.

XRT data were downloaded from the Swift Data Center at NASA/GSFC
(level 1 data products). 
They were then calibrated, filtered, and screened using the \texttt{XRTDAS}
software package (v.2.3) developed at the ASI Science Data Center (ASDC)  
to produce cleaned photon list 
files\footnote{\texttt{http://swift.gsfc.nasa.gov/docs/swift/analysis/xrt\_\\swguide\_v1\_2.pdf}}. 
The temperature of the CCD was acceptably below $-50$~\degr{}C
for the whole observation set.
The total exposure times after all the cleaning procedures were 8.2~s, 9.6~s, 
and 368\,815~s for data accumulated in LR, WT, and PC mode, respectively. 

For both the spectral and timing analyses we used standard grade selections:
$0-12$ for PC mode, $0-5$ for LR mode, and $0-2$ for WT
mode. However, data in WT mode had {\rm insufficient} statistics
to allow for detailed spectral modeling and were used only 
in the light curve analysis. 
Ancillary response files for PC and LR spectra were generated through the
standard {\tt xrtmkarf} task (v0.5.1) using the response files 
{\tt swxpc0to12\_20010101v007.rmf} and {\tt swxpd0to5\_20010101v007.rmf} 
from {\tt CALDB} (2006-01-04 release). 

In the timing analysis, XRT times are referred to the XRF~050416A BAT trigger time 
$T= 2005$ Apr 16.461626 UT (2005 Apr 16, 11:04:44.5 UT). 

\section{XRT DATA ANALYSIS}
\label{xrtanalysis}

\subsection{SPATIAL ANALYSIS}
\label{spatial}

\begin{figure*}[ht]
\centerline{\includegraphics[width=15cm,angle=0]{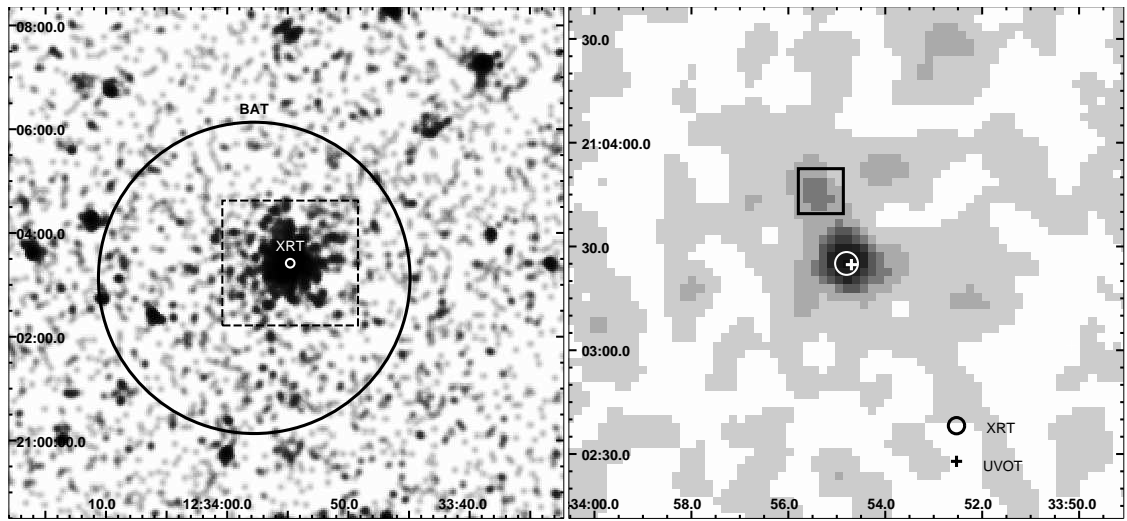}}
\caption{Left: Photon Counting mode XRT image of the first and second observations 
(92 ks exposure), smoothed with a Gaussian filter with a 3 pixel standard deviation,
showing the XRT and BAT error circles 
{\rm (\citealp{moretti05}, \citealp{saka2}, respectively)}. 
The dashed box identifies the area enlarged in the right panel.
{\rm The UVOT afterglow position is not marked on this image because 
it is indistinguishable from the XRT one.}
Right: Cumulative image of the follow-up observations from 3 to 13 
(152 ks exposure; Table~\ref{log}), showing the XRT error circle,
{\rm the UVOT position as given in \citet{scha1} (cross) and the position 
of the second source detected within the BAT error circle (square).}}
\label{fov}
\end{figure*}

Figure~\ref{fov} 
(left panel) shows the XRT image accumulated in PC mode with a 
$0.2-10$~keV energy selection during the first and second 
observations, together with the BAT {\rm and XRT error circles}. 
The central portion of this field is expanded in the right panel 
of Fig.~\ref{fov}, in which we show the cumulative image 
of the follow-up observations \# 3 to 13.
Two sources are visible within the BAT error circle. The brighter one
is coincident {\rm with the position of the optical counterpart 
as derived by UVOT (cross point)} and it is clearly fading with time.
Therefore we identify it as the afterglow of XRF~050416A. The fainter
source lies 21$\farcs$6 away from the UVOT afterglow \citep{scha1}
and it does not show any significant evidence of intensity
variations during the XRT observations. Its count rate,
as determined from the sum of the follow-up observations \# 3 to 13
(chosen to minimize the contamination from the afterglow emission)
is $(3.3\pm 0.7)\times10^{-4}$~count~s$^{-1}$.

The afterglow position derived with {\tt xrtcentroid} (v0.2.7) 
is RA$_{\rm J2000} = 12^{\rm h} 33^{\rm m} 54\fs62$, 
Dec$_{\rm J2000} = +21\degr 03\arcmin 27\farcs7$,
with an uncertainty of 3$\farcs$3. 
This position takes into account the correction for the
misalignment between the telescope and the satellite optical axis
\citep{moretti05}.
The XRT boresight corrected coordinates are 
45$\farcs$0 from the BAT position \citep{saka2} 
and 0$\farcs$9 from the optical counterpart \citep{scha1}.

\begin{figure*}[ht]
    \epsscale{0.8}
\centerline{\includegraphics[width=14cm,angle=+90]{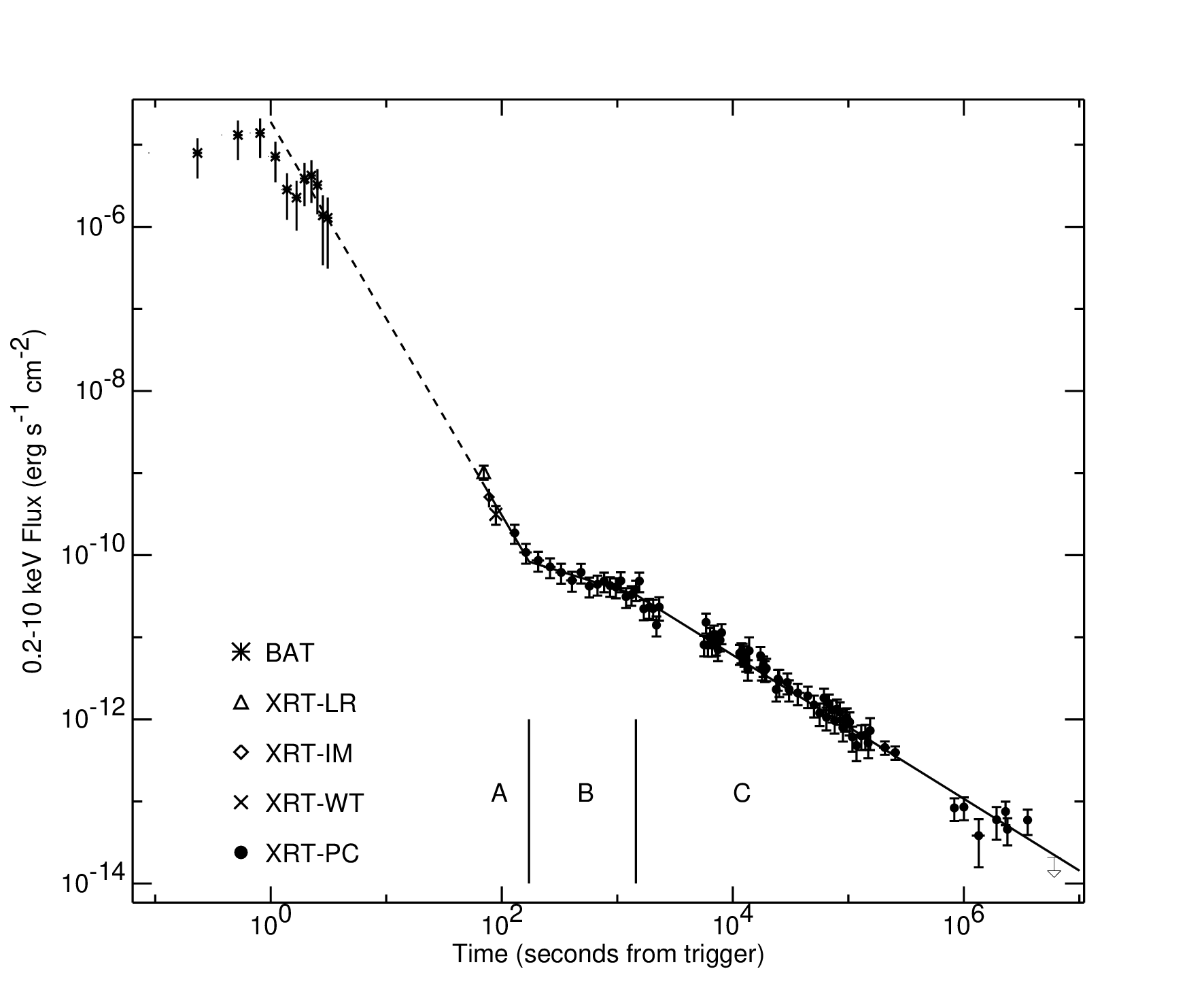}}
    \caption{BAT and XRT light curves of XRF~050416A. 
The XRT count rate ($0.2-10$~keV) was converted into flux units 
by applying a conversion factor derived from the spectral analysis. 
The solid line represents the best-fit model with the doubly-broken power law.
The dashed line is the extrapolation of the XRT best fit model prior 
to the first XRT observation. 
The BAT light curve was extrapolated into the $0.2-10$~keV XRT energy band 
by converting the $15-150$~keV BAT count rate 
through the single power law best fit model with photon index $\Gamma=3.1$ 
presented in \citet{saka4}. 
}
\label{lc}
\end{figure*}

\subsection{TIMING ANALYSIS}  
\label{timing}

The X-ray emission from the counterpart of XRF~050416A was detected for the first time 
in the settling data (i.e. those data collected during the satellite slew, when the XRT
pointing direction was less than 10\arcmin{} off the target position).
We have a total of 8.2~s of data in which the source is significantly
detected before the ``official'' beginning of the pointed XRT observation. 
In these data the maximum offset of the source is lower than 3$\arcmin$ and 
no vignetting correction was thus required.
As LR is a non-imaging mode, the background has been extracted from
pointed LR data taken at the beginning of {\rm an} orbit,
$1.5\times10^5$~s after the burst trigger, when, as we
know from the PC data (see below), the afterglow emission had faded to
less than 1\% of the initial value. This background subtraction allows us to
correctly account for the emission of the serendipitous non-transient sources 
in the field of view in addition to the instrumental and cosmic X-ray background.

After settling, a single exposure (2.5~s) image mode frame was taken,
which officially marks the beginning of pointed observation. 
We determined the {\rm total amount of charge above the background 
inside a circular region of 30 pixel radius centered at the source position 
in raw telemetry units (i.e. Data Number or DN units)}. 
We obtained  278 DN above a background of 0.5 DN. 
The DN value was converted to count rate by evaluating 
the mean energy of the LR spectrum in the $0.2-10$~keV band (1490~eV). 
Given a mean energy per DN of $\sim$79~eV in the low gain imaging mode we
calculated a rate of $5.9\pm1.5$~count~s$^{-1}$.

WT data were extracted in a rectangular region 40 pixels wide 
along the image strip, which includes about 98\% of the 
point spread function (PSF). 
The background level was extracted from a rectangular 
region of the same extension, far from the source and affected by minimal 
contamination from other sources in the field. The rapid decay of the source 
and the high background level in WT mode allow us to have a significant 
detection of the XRF~050416A only in the first 8 frames (9.6~s) of the 
WT data set.

Given the pile-up effect in the first part of the observation, the 
presence of the second source and the weakness of the afterglow after 
the first follow-up observation, the PC data require different extraction regions 
for different rate levels, in order to optimize the S/N ratio.\\
\indent {\it First orbit.} 
The intensity of the source during the first orbit of the first observation 
(Table~\ref{log}) was high enough to cause pile-up in the PC frames. 
In order to correct for this effect, we extracted counts from an annular region 
with an outer radius of 30 pixels (70$\farcs$8) and an inner radius of 3 pixels 
(7$\farcs$1). Such a region includes about 52\% of the PSF.
The optimal inner radius was evaluated by comparing 
the analytical PSF with the profile extracted in the first 2000~s 
of observation. {\rm The excluded region corresponds to pixels deviating 
more than one sigma from the best fit of the differential PSF wings 
(i.e. the fit performed on data from 7 pixels outwards)}.
The light curve was corrected for the PSF fraction loss.\\
\indent {\it First observation (excluding the first orbit) and second
observation.} 
In the following 28 orbits of the first observation and throughout the second 
observation the intensity of the afterglow was lower than 0.1~count~s$^{-1}$ 
and the pile-up {\rm was} negligible. The data were extracted from the entire 
circular region of 30 pixels radius in order to have the maximum 
available statistics, particularly important in the last part of the afterglow  
decay. This circle encloses 93\% of the PSF. Both here and in the 
previous case, the contribution of the second, fainter source detected within 
the BAT error box is negligible with respect to the afterglow intensity. \\
\indent {\it Observations 3 to 13.}
The afterglow had faded to a count rate comparable to that of 
the serendipitous source. In order to avoid significant 
{\rm (i.e. greater than 10\%)} contamination, we reduced the extraction 
radius to 6.5 pixels (15$\arcsec$). 
Such a region includes about 71\% of the source PSF and the possible contamination 
from the nearby source within this region amounts to about 12\% of its PSF. 
Given the faintness of the afterglow in this final part of the light-curve, 
this choice also improves the S/N ratio. 

The background level for the PC data was extracted in an annular region with 
an inner radius of 40 pixels and an outer radius of 150 pixels centered 
at the source position.
{\rm To eliminate contributions from faint sources in the background region,
we produced a 380~ks image by summing all of the observations, and searched
this image for faint sources within the background annulus. In addition
to the serendipitous source shown in  Fig. \ref{fov}, 20 other sources
were found with S/N ratio higher than 3, all of which were located more
than 50 pixels from the afterglow. The contributions from these sources
were excluded from the background region.}

Data were binned in order to have a S/N ratio higher than 3. The source was not 
detectable after observation \# 13; data from observations 14-19 were 
summed together and provide a single 3~$\sigma$ upper limit value of  $2.3\times 
10^{-4}$~count~s$^{-1}$.
Figure~\ref{lc} shows the background-subtracted light curve in the $0.2-10$~keV 
energy band. The source is clearly fading with time. 

The XRT light curve decay is not consistent with a single power law 
($\chi_{\rm red}^{2}=1.43$, with 80 degrees of freedom, d.o.f.). 
A broken power law: 
\begin{displaymath} 
F(t)= \left\{ \begin{array}{ll} 
K~t^{-\alpha_{\rm A}}  & \textrm{ {\rm for}~ $t < T_{\rm b}$ }\\
K~T_{\rm b}^{\alpha_{\rm B}-\alpha_{\rm A}} t^{-\alpha_{\rm B}} & \textrm{ {\rm for}~ $t \ge T_{\rm b}$ }\\
\end{array} \right. 
\end{displaymath}
(where $\alpha_{\rm A}$ and $\alpha_{\rm B}$ are the  power-law slopes before
and after the break time $T_{\rm b}$, respectively)
improves the fit, giving $\chi_{\rm red}^{2}=1.20$ (78  d.o.f.). 
However, the residuals show a systematic trend. 
Adding a second break to the model: 
\begin{displaymath}
F(t)= \left\{ \begin{array}{ll}
K~t^{-\alpha_{\rm A}}  & \textrm{ {\rm for}~ $t < T_{\rm b,1}$ }\\
K~T_{\rm b,1}^{\alpha_{\rm B}-\alpha_{\rm A}}~t^{-\alpha_{\rm B}} 
& \textrm{ {\rm for}~ $T_{\rm b,1} \le t < T_{\rm b,2}$ } \\
K~T_{\rm b,1}^{\alpha_{\rm B}-\alpha_{\rm A}}~T_{\rm b,2}^{\alpha_{\rm C}-\alpha_{\rm B}}~t^{-\alpha_{\rm C}}
& \textrm{ {\rm for}~ $t \ge T_{\rm b,2}$ } \\
\end{array} \right.
\end{displaymath}
the fit further improves yielding $\chi_{\rm red}^{2}=0.81$ 
(76 d.o.f.). The F-test for this model vs. the simple broken power-law 
gives a chance probability of $1.2 \times 10^{-7}$. 
This last model reveals the presence of two  breaks
at $172 \pm 36$ s  and at (1.450 $\pm$ 0.013) $\times 10^3$ s.
Table~\ref{lcfit} shows the best fit results obtained with the three models.
We also tried to fit the light curve with a single or broken 
power law allowing the reference time $t_0$ to be a free parameter.
For the simple power law and the single break broken power-law
$\chi^2_{\rm red}$ improved to 1.40 (79 d.o.f.) and 1.02 (77 d.o.f.), 
respectively, but the best-fit model is not able to account for the 
initial part of the light curve decay and the resulting $t_0$ 
is before the burst onset 
($t_0=-201 \pm 156$~s and $t_0 =-153 \pm 34$~s, respectively). 
For the doubly-broken power law the improvement is not significant 
and the value of $t_0$ is marginally consistent with the burst
trigger time.

In Fig.~\ref{lc} we plot the data as well as the best fit model
obtained with the doubly-broken power law, 
with $t_0 = 0$. We also show the model extrapolation back to the
time of the trigger.
Hereafter, we will refer to the time intervals $t < T_{\rm b,1}$,
$T_{\rm b,1} < t < T_{\rm b,2}$, and $t > T_{\rm b,2}$ as
phases `A', `B', and `C', respectively.

{\rm Note that the second observation starts after the second
break in the light curve (see fifth column in Table~\ref{log}),
and all other observations contribute only to phase C.
Observations from 3 to 13 correspond to the last seven points of
the light curve shown in Fig.~\ref{lc}, and observations 
from 14 to 19 give the final upper limit.} 

\begin{deluxetable}{cccc}
\tablecolumns{5}
\tabletypesize{\normalsize}
\tablecaption{XRF~050416A light curve best fit parameters}
\tablewidth{0pt}
\tablehead{
\colhead{Parameter} &
\colhead{Single  PL}&
\colhead{Broken PL}&
\colhead{Doubly-broken PL} 
}
\startdata
$\alpha_{\rm A}$              & 0.82 $\pm$ 0.02 & 3.28  $\pm$ 0.04 & 2.4 $\pm$0.5     \\ 
$T_{\rm b,1}$ (s)             &  --             & 103 $\pm$ 9      & 172 $\pm$ 36     \\ 
$\alpha_{\rm B}$              &  --             & 0.81 $\pm$ 0.02  & 0.44 $\pm$ 0.13  \\ 
$T_{\rm b,2}$ (10$^3$ s)      &  --             &  --              & 1.450 $\pm$ 0.013\\ 
$\alpha_{\rm C}$              &  --             &  --              & 0.88 $\pm$ 0.02  \\ 
$\chi_{\rm red}^{2}$ (d.o.f.) &  1.43 (80)      &  1.20 (78)       & 0.81 (76)        \\ 
\enddata
\label{lcfit}
\tablecomments{$\alpha_{\rm A}$, $\alpha_{\rm B}$, and $\alpha_{\rm C}$ are the decay slopes for
the distinct phases of the light curve (see \S~\ref{timing}). 
$T_{\rm b,1}$ and $T_{\rm b,2}$
are the epochs at which the decay slope changes,
measured from the XRF onset.
The IM and LR points have been included in the fits.}
\end{deluxetable}

\begin{figure}[h]
\centerline{\includegraphics[width=11cm,angle=-90]{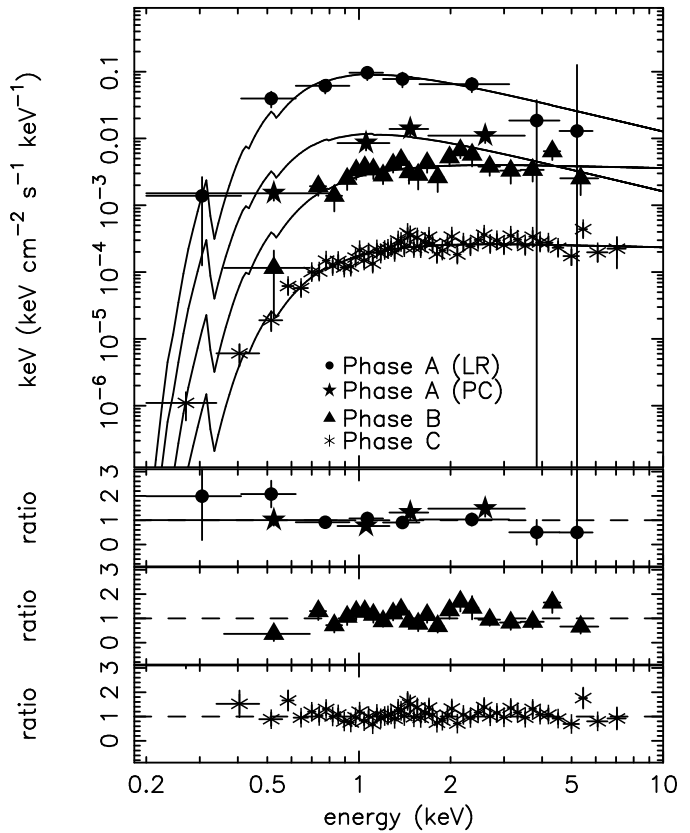}}
    \caption{
Phase A, B, and C spectra converted in the $E^2 N(E)$ (or $\nu F_{\nu}$) 
representation together with their best fit absorbed power-law models. The panels 
showing residuals refer to phase A, B, and C, respectively from top to bottom.
 }
\label{spectra}
\end{figure}

\subsection{SPECTRAL ANALYSIS} 
\label{spectral}

Detailed spectral analysis could be performed only for the LR data 
in settling mode acquired before the start of the first observation
and on the PC data of the first and second observations; 
the following 11 observations did not add significant statistics to the
spectra, thus they were not included in the analysis. The same holds for the WT
data of the first observation consisting of only 31 (background-subtracted) 
photons. 

Given the presence of three different phases in the XRF~050416A light curve 
decay, we first checked for spectral variability during the afterglow evolution. 
The LR spectrum and background were extracted as described in 
\S~\ref{timing}. 
The two PC spectra for the phases A and B in Fig.~\ref{lc} 
were extracted {\rm from the first observation with} the same 
annular region used for the timing analysis, 
while the PC spectrum for the phase C was extracted from 
{\rm the remaining part of the first observation and the 
second observation (i.e. up to $2.8 \times 10^5$ s) using} 
the circular region of 30 pixel radius. 
The PC background spectra were extracted from the same region 
as for the timing analysis. 

We fitted the spectra from phases A, B, and C separately with
absorbed power laws. For phase A, the LR and PC spectra were fitted
together leaving the normalization parameters free to take into
account the different rate level due to the light curve decay.
The four spectra are shown in Fig.~\ref{spectra} together with their
absorbed power law best fits.

The best fit results show evidence for spectral variation among phases:
the emission in phase A is significantly softer than in the phases B and
C. In the latter two phases the best fit photon indices were consistent
within the errors. 
Spectra extracted from phases B and C were, therefore, summed together 
and their ancillary response files, obtained from different extraction regions,
were weighted according to the relative exposure.
The fit of the resulting spectrum gave an absorption column density of
$(2.6^{+0.4}_{-0.3})\times10^{21}$~cm$^{-2}$ and a photon index 
$\Gamma=2.04_{-0.05}^{+0.11}$, with a reduced $\chi^2$ of 1.2 (83 d.o.f.). 
The observed column density is significantly larger than the Galactic value 
($0.21\times10^{21}$~cm$^{-2}$). 
We therefore checked for intrinsic absorption in the host galaxy by adding
a redshifted absorption component (\texttt{zwabs} model in XSPEC v11.3.1)
with the redshift fixed to 0.6535 \citep{cenko3} and the Galactic absorption
column fixed to $0.21\times10^{21}$~cm$^{-2}$.  
The fit gave a value of $6.8^{+1.0}_{-1.2} \times10^{21}$~cm$^{-2}$ for 
the additional column density with an improvement of the reduced 
$\chi^2$ to 1.0 (83 d.o.f.). 
The phase A spectra, that also showed a column density significantly
higher than the Galactic value, were fitted again with the
addition of a redshifted absorption component. Because of the low statistics
of the phase A data, the column density was kept fixed to the value 
obtained from the phase B$+$C spectral fit and the best fit photon index
was $\Gamma = 3.0^{+0.3}_{-0.4}$ ($\chi^2_{\rm red} =0.6$; 5 d.o.f.).
Table~\ref{tspec} shows the final results of the spectral analysis.

\begin{deluxetable*}{lll}[ht]
\tablecolumns{4}
\tablecaption{XRF~050416A XRT spectral fit results}
\tablewidth{0pt}
\tablehead{
\colhead{Parameter} &
\colhead{Phase A}&
\colhead{Phase B+C}
}
\startdata
Galactic column density (\NHunits)                         & 0.21 (frozen)                & 0.21 (frozen) \\
Host column density (\NHunits)                             & 6.8  (frozen)                & $6.8^{+1.0}_{-1.2}$ \\
Photon index                                               & $3.0^{+0.3}_{-0.4}$          & $2.04^{+0.11}_{-0.05} $ \\
$N$ (photons keV$^{-1}$ cm$^{-2}$ s$^{-1}$ at 1 keV)       & $0.14\pm0.03$                & $(2.7\pm 0.3)\times 10^{-4}$ \\
$0.2-10$~keV flux (\flux)                                  & ($1.9\pm0.1)\times 10^{-10}$ & $(1.7 \pm 0.2)\times 10^{-12}$ \\
$0.2-10$~keV luminosity (erg s$^{-1}$)                     & $5.7 \times 10^{47}$         & $3.1 \times 10^{45}$  \\
$\chi_{\rm red}^{2}$ (d.o.f.)                              & 0.6 (5)                      & 1.0 (83) \\ 
\enddata
\label{tspec}
\tablecomments{The intrinsic column density for the LR data was held fixed to
the best fit value found from the PC spectrum. The (isotropic) luminosity was
calculated for a redshift $z = 0.6535$, 
with $H_0$=70~km~s$^{-1}$~Mpc$^{-1}$, $\Omega_{\rm m}$=0.3, and $\Omega_{\Lambda}$=0.7. 
Unabsorbed fluxes and luminosities reported for the PC data are averaged 
over long time intervals: accurate instantaneous values 
for the unabsorbed flux can be derived from Fig.~\ref{lc}.}
\end{deluxetable*}

\section{DISCUSSION}
\label{discussion}

We have presented a detailed analysis of the X-ray afterglow of XRF~050416A. 
The prompt emission of this burst lasts $\sim 2.5$~s and 
is characterized by a first peak followed by a second much weaker one. 
The event belongs to the short tail of the long GRB population
\citep{kouve93}. It could also be consistent with the third class
of bursts identified by \citet{mukhe98} {\rm through multivariate analysis
on the BATSE catalog. This class consists of bursts of intermediate duration 
and fluence as compared to the standard classes of short/faint/hard 
and long/bright/soft GRBs.}
The average energy distribution of the prompt emission is soft
(well fitted by a power law with $\beta=\Gamma-1= 2.1\pm0.2$ at 
the 90\% confidence level or by a Band model with $\alpha_{\rm Band} \equiv -1$, 
$E_{\rm p}=15.6_{-2.7}^{+2.3}$~keV, and $\beta_{\rm Band}< -3.4$ 
at the 68\% confidence level) 
with significant evidence of hard to soft evolution within each peak
\citep{saka4}.
The gamma-ray spectral distribution of this burst, 
with fluence in the X-ray energy band $2-30$~keV larger
than the fluence in $30-400$~keV band, 
classifies it as an X-ray flash \citep{lamb,saka3}.

XRT monitored the XRF~050416A X-ray emission from $\sim 64.5$~s 
up to 74~d after the BAT trigger. 
XRF afterglows have been rarely detected in the past 
(XRF~011030, XRF~020427: \citealp{bloom03b,levan05a};
XRF~030723: \citealp{butler05}; XRF~040701: \citealp{fox04};
XRF~050215B: \citealp{levan05b}; XRF~050315: \citealp{vaughan05}; 
XRF~050406: \citealp{bur05b,romano05}; XRF~050824: \citealp{krimm05}). 
The exceptionally long observational campaign 
of XRF~050416A has provided us with a unique data set 
and allowed one of the most accurate spectral 
and timing analyses ever performed for an XRF afterglow. 

The XRF~050416A light curve of the first $10^5$~s 
after the trigger (Fig.~\ref{lc}) is fairly smooth and similar in shape to
other Swift-detected XRF and GRB afterglows like 
XRF~050315 \citep{vaughan05} or GRB~050319 (\citealp{cusumano05b}; 
see also \citealp{nousek05,chinca05,obrien05}).
It shows evidence of three different phases (A, B, and C according to
\S~\ref{timing}), each of them characterized by a distinct 
decay slope (see Fig.~\ref{lc} and Table \ref{lcfit}).
At the beginning of the XRT observation the light curve shows
a steep decay ($\alpha_{\rm A} \sim 2.4$), followed by a short 
flat phase ($\alpha_{\rm B} \sim 0.44$) and then by a third long-lasting 
phase with a more rapid intensity decline ($\alpha_{\rm C} \sim 0.9$).
We also found that the late extrapolation of the phase C decay
is consistent with the flux upper limit measured $65-74$~d
after the prompt emission (observations \# $14-19$). 
There is no evidence of X-ray flares as seen in
XRF~050406 \citep{romano05},
GRB~050502B \citep{bur05b,falcone05}, 
GRB~050607 \citep{pagani05}, and many other events.

The XRT spectra show significant excess absorption in the rest 
frame of XRF~050416A ($N_{\rm H} \sim 6.8 \times 10^{21}$~cm$^{-2}$)
and an energy distribution significantly softer in phase A 
($\beta = 2.0 \pm 0.4$)
than in phases B and C, which have consistent spectral slope
with a weighted average energy index $\beta = 1.04 \pm 0.05$.
This may indicate a different emission process acting during the
initial phase of the XRF~050416A light curve.

\subsection{PHASE A}
\label{discussion:phaseA}

In the internal/external shock scenario \citep{rees94}, the tails of GRB peaks
are expected to be caused by the ``high latitude effect''
\citep{kumar00,dermer04}. If a relativistic shell of matter suddenly stops
shining, a distant observer receives photons emitted from increasing off-axis
angles at later times due to the longer travel path. 
When the observed frequency is above the synchrotron cooling frequency
(which is usually the case in the X-ray band), 
then the decay index expected for the observed light curve is $\alpha = 2+\beta$, 
where $\beta$ is the spectral index measured during the decay. 
The decay slope of phase A ($\alpha_{\rm A} = 2.4 \pm 0.5$) is definitely lower 
than the value expected for high-latitude emission ($\sim 4.0 \pm 0.4$). 
However, the high-latitude effect only provides an upper limit to the decay slope, 
since it assumes that the shell emission stops abruptly after the initial pulse. 
Residual emission may still be present from the shocked shells, so that 
slower decline rates are possible. The high-latitude emission would in this case 
contribute only a small fraction of the overall radiation. Another possibility
is that the X-ray band was below the cooling frequency: in this case the expected
decay slope of high latitude radiation would naturally be shallower than 2+$\beta$.

In this scenario the first break in the X-ray light curve would be due to the
emergence of the afterglow light {\rm after fading of} the prompt emission (see
Sect.~\ref{discussion:phasesBC}). The forward shock may therefore contaminate
the tail emission. This component would however make the spectrum harder. Since
most of the phase A counts come from the LR data, that correspond to the very
first point in the X-ray light curve, we expect the afterglow to provide little
contribution ($\sim 10$\%) at this time.

\subsection{PHASES B AND C}
\label{discussion:phasesBC}

According to \citet{zhang05}, the standard interpretation of 
the flat decay slope during phase B
and of the second temporal break in the afterglow
involves refreshed shocks \citep{refreshedb}. 
In the initial stages of the fireball evolution the forward shock,
which emission produces the X-ray afterglow, 
may be continuously refreshed with the injection of additional energy. 
This could happen either because the central engine still emits 
continuously \citep{dailu98,zhangmes2001,dai04}, 
or because slower shells emitted at the burst time catch up the fireball
which has already decelerated
\citep{rees98,pana98,kumarpiran00,refreshedb,zhangmes2002}.
Within this scenario, a flat decay of the afterglow
is expected as the refreshed forward shock decelerates less rapidly 
than in the standard case. 
A transition to the standard afterglow evolution (i.e. a break) 
with no remarkable spectral changes is also expected when the
additional energy supply ends.
According to the \citet{zhang05} analysis, the various refreshing mechanisms 
can be characterized by an effective index $q < 1$ such that
a decay slope $\alpha_{\rm inj}=(1+q/2)\beta+q-1$ is expected
for the afterglow light curve until injection stops. 
For XRF~050416A, with $\beta\sim1.0$ and 
$\alpha_{\rm inj} = \alpha_{\rm B} \sim 0.44$, 
we can derive $q\sim0.3$. 

Note that the phase C decay slope and spectral index are 
marginally consistent with $\alpha_{\rm C}=(3p-2)/4$ and $\beta_{\rm C}=p/2$
for $p \sim 2$. This is expected for a fireball propagating 
in a uniform interstellar medium or in a stellar wind environment
when the synchrotron cooling frequency is below the X-ray region, 
and before the jet break \citep{sapina98,chevalier00}.
Since phase C remarkably continues uninterrupted 
until the last XRT detection 42 days after the burst,
this interpretation implies the absence of both 
a cooling break (expected in case of wind environment) 
or a jet break in the X-ray afterglow.
Wind and magnetic field parameters can be easily adjusted
to delay the cooling break after the end of the observational
campaign of XRF~050416A, thus the circumburst environment 
cannot be distinguished based on X-ray data alone.

\begin{figure*}
\epsscale{0.6}
\centerline{\includegraphics[width=14cm,angle=+90]{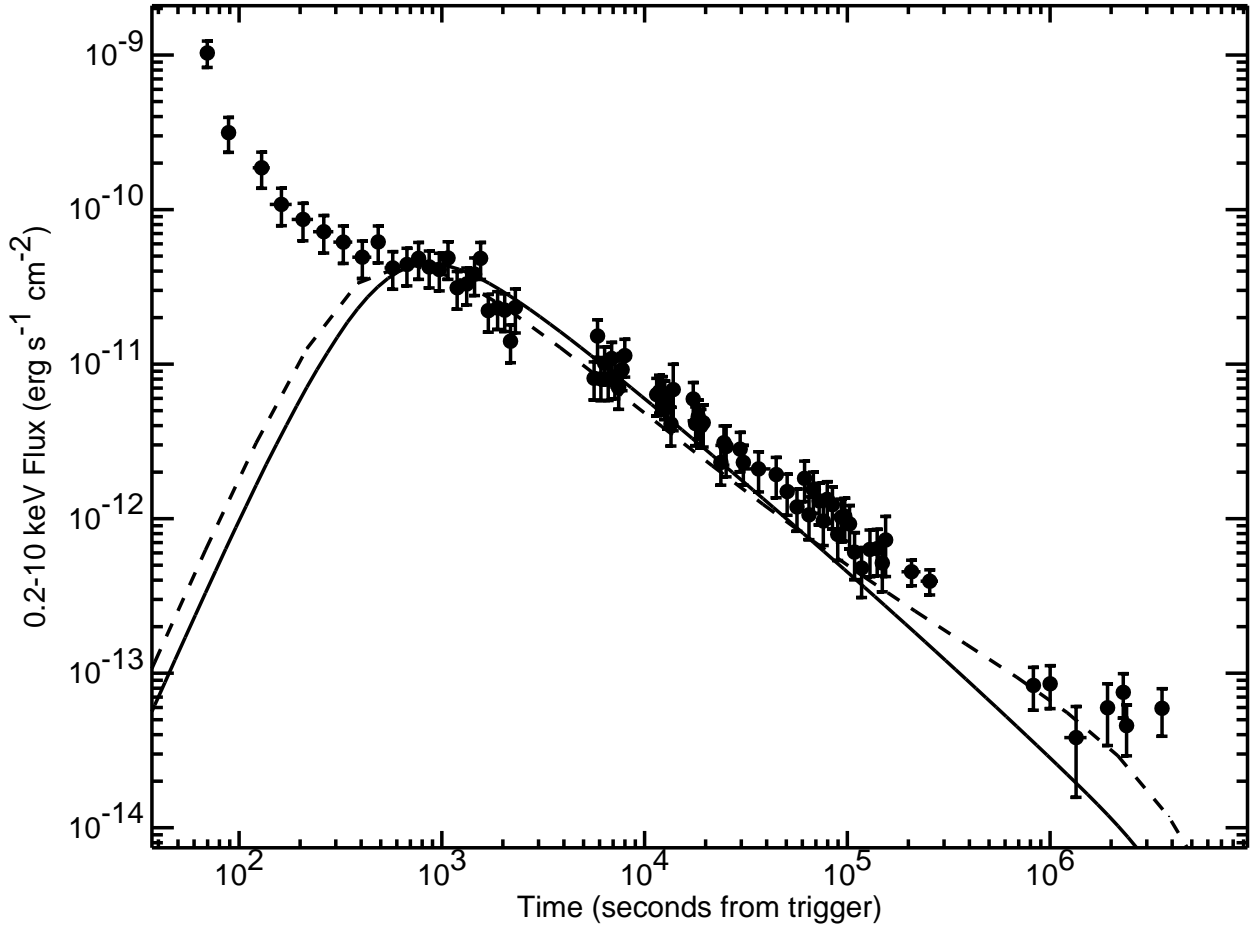}}
\caption{Two model light curves plotted on the XRT data.
The solid line shows the results for a uniform jet viewed
on-axis ($\zeta=0^\circ$) with opening angle $\theta_{\rm jet} = 17^\circ$, 
$\Gamma_{\rm blk}=80$, and $E=2\times10^{51}$ erg. 
We have assumed $n=0.1$ cm$^{-3}$, and $p=2.08$. 
The dashed line shows the results for a structured jet 
with a Gaussian distribution of outflow energy 
and with a uniform distribution of $\Gamma_{\rm blk}$,
viewed off-axis ($\zeta=10^\circ$). 
The 1$\sigma$ width of the Gaussian was $4^\circ$. 
The other parameters are the same as in the uniform case.
\label{theory}}
\end{figure*}

\begin{figure}
\epsscale{0.6}
\centerline{\includegraphics[width=6cm,angle=-90]{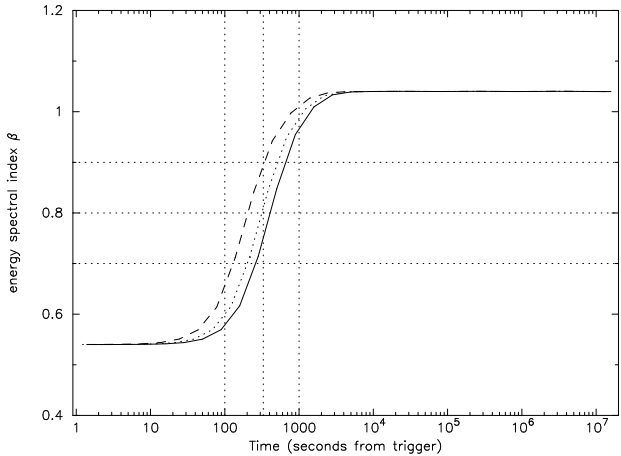}}
\caption{The temporal evolution of the energy spectral index $\beta$
calculated for models with $n=0.1$ cm$^{-3}$, $E=2\times10^{51}$ erg
and three different values of the bulk Lorentz factor $\Gamma_{\rm blk}=100,
80, 70$ (dashed, dotted, and solid). The index was calculated as an 
arithmetic average of 5 values uniformly distributed within the XRT 
energy band.
\label{beta_evolv}}
\end{figure}

\subsection{JET MODELS}
 
The long and well sampled X-ray light curve of XRF~050416A is
one of the best suited for testing more realistic jet models.
The leading jet models for XRFs include an on-beam uniform jet
with a very wide opening angle \citep{lamb}, 
an off-beam uniform jet with the line of sight outside the jet edge 
\citep{yamazaki03}, a structured Gaussian-like jet with the viewing angle 
outside the bright Gaussian core \citep{zhang04a}, and a two-component 
jet with the line of sight on the less energetic wider beam 
such as the cocoon surrounding a collapsar jet \citep{zhang04b}. 
The segment C displays a ``normal'' afterglow decay without significant 
features. This rules out the off-beam uniform jet which predicts 
an initial fast rise and a rapid decay (e.g.  \citealp{granot05}). 
The lack of a rebrightening feature also greatly constrains 
the two-component jet. We therefore only focus on two 
possibilities, i.e. an on-beam uniform jet and an off-beam 
structured jet.

In this section we describe preliminary results of our efforts to model
the light curve in the phase C through numerical simulations.
We model the synchrotron radiation powered by the external shock 
propagating in a uniform medium with proton number density $n$. 
Our three dimensional code, which was developed by J. Dyks, can
calculate afterglow emission of an arbitrary axially symmetric jet
observed at any viewing angle. The code has been used to
model the curvature effect in structured jets \citep{dyks05}
and more generally the GRB afterglow light curves (described in
Zhang et al. 2005). The code takes into account all kinematic effects
that affect the observed flux (e.g.~Doppler boost, propagation time delays)
in the way described by \citet{s03}.
The radial dynamics of the outflow is followed well into the non-relativistic
regime using the equations of \citet{huang00}. Sideways expansion is 
neglected. The evolution of the electron energy spectrum is followed with 
the simplified analytical method \citep{fww04}, except for the 
top-hat (i.e. uniform jet) case when the low run time has allowed 
exact integration of the continuity equation \citep{msb00}. 
In the calculations discussed below, we have used the exact
spectrally derived value of $p=2\beta_{\rm C}=2.08$, which
assumes the cooling frequency was below the X-ray band.

Light curves calculated for a uniform jet viewed on-beam are notably steeper 
than the data (the predicted slope is $\alpha=1.1$, to be compared with the
observed value $\alpha_{\rm C} = 0.88\pm0.02$).
The one shown in Fig.~\ref{theory} (solid line) has been calculated for
the half opening angle of the uniform jet $\theta_{\rm jet}=17^\circ$,
the viewing angle $\zeta=0$, the total explosion energy of the two-sided 
outflow $E=2\times10^{51}$ erg, the bulk Lorentz factor $\Gamma_{\rm blk}=80$,
{\rm the proton number density of the external medium} 
$n=0.1$~cm$^{-3}$, and the electron and magnetic field energy equipartition
parameters, $\epsilon_{\rm e}$ and $\epsilon_{\rm B}$, equal to 0.1 and 0.01, 
respectively.
A ``contrived" version of structured jets, with uniform distribution 
of $\Gamma_{\rm blk}$ as a function of $\theta$, but with the explosion 
energy $\epsilon(\theta)$ increasing towards the jet axis can produce 
light curves that are less inconsistent with the data 
(dashed line in Fig.~\ref{theory}).
This is thanks to the well known flattening that appears 
near the jet break time for large viewing angles $\zeta > \theta_{\rm core}$, 
where $\theta_{\rm core}$ is the half opening angle of the most energetic 
central part of the outflow. However, the light curves typically have 
a slightly concave shape that is not observed {\rm in} phase C.
The dashed line in Fig.~\ref{theory} is for a structured jet with 
the uniform $\Gamma_{\rm blk}$, and a Gaussian profile of $\epsilon(\theta)$
with standard deviation $\sigma=\theta_{\rm core}=4^\circ$ and $\zeta=10^\circ$.
The other parameters are the same as in the top-hat case.

Standard theoretical parameter values of $\Gamma_{\rm blk}=$100, $n=1$~cm$^{-3}$ 
and an assumption of $E=2 \times 10^{51}$~erg result in models in which 
the cooling frequency is below the X-ray band throughout phase B, 
with the result that the model slope is steeper than the observed 
slope in that phase. The flatter slope of phase B could be interpreted 
as a ``cooling break'' if the cooling  frequency begins above the 
XRT energy band and crosses it at the phase B / phase C transition.  
This would require lower $\Gamma_{\rm blk} \sim 70-80$, 
lower $n \simeq 0.1$~cm$^{-3}$, or larger explosion energies 
$E \gtrsim 10^{52}$ erg.  The low value of $n$ and large $E$ 
could also help explain the late jet break time without requiring 
a large solid angle for the uniform outflow.  
However, the presence of a cooling break at the phase B / phase C 
transition should be associated with a strong spectral change, 
which is not observed.  In fact, the modeled energy spectral index 
in this case changes by 0.5 at this transition (see Fig.~\ref{beta_evolv}), 
in contrast to the observations.  The hard model spectrum in phase B 
could in principle be compensated for a contribution from the softer, 
prompt emission, but this seems contrived, as it would have to balance 
perfectly to produce no change in the observed spectral index.  
The interpretation given above for phase B in terms of refreshed shocks 
seems more plausible.

\section{CONCLUSIONS}
\label{section::conclusions}

{\rm The steep initial phase of the X-ray light curve of XRF~050416A
can be interpreted as the tail of the prompt emission.}
The rest of the light curve shows evidence of forward shock 
refreshing mechanisms acting up to about 1450~s since the trigger 
and then an uninterrupted decay with no signature of jet breaks 
up to at least 42 days.

The overall phenomenology of XRF~050416A and other 
Swift detected XRFs provides evidence that both GRBs and XRFs 
arise from the same phenomenon \citep{lamb,saka3,dalessio05}, 
but the characteristics of XRF~050416A deviate somewhat 
from model expectations.  Our preliminary numerical simulations 
confirm that the late-time decay of XRT~050416A is slightly 
flatter than expected for on-axis viewing of a uniform jet.  
Modeling of off-axis viewing of a structured jet with a uniform 
bulk Lorentz factor and a Gaussian internal energy distribution 
also could not reproduce the X-ray light curve satisfactorily 
for reasonable values of the model parameters.  More realistic 
afterglow models should be investigated in detail to understand 
these effects and to fully explain the broad range of peak energies 
observed from XRFs through X-Ray Rich bursts to classical GRBs.

Swift XRT results are showing that jet breaks are rare for both 
XRFs and GRBs (XRF~050315, \citealp{vaughan05};
GRB~050318, \citealp{perri05}; GRB~050505, \citealp{hurkett06};
GRB~050525A, \citealp{blustin05}). 
Because of its relative slow decay rate and its exceptionally 
long X-ray light curve, XRF~050416A provides a particularly 
stringent case for understanding jet collimation and structure 
in GRBs.  
{\rm A jet opening angle $\theta_{\rm jet} \simgt 28^{\circ}$ 
can be estimated through Eq.~1 in \citet{sari99} 
assuming  $t_{\rm jet} \simgt 42$~d
(i.e. the time of the last detection in the XRT light curve),
an isotropic energy $E_{\rm iso} \simlt 1.2 \times 10^{52}$~erg
(obtained from the BAT power law fit) and standard values of
the proton number density of the external medium $n=3$
and the radiative conversion efficiency of the burst $\eta_{\gamma}=0.2$.}
This opening angle far exceeds the jet angle of 
$5^{\circ} - 10^{\circ}$ generally considered to be typical for GRBs.  
Whether this is related to the soft nature of this event requires 
a larger sample of jet breaks for both GRBs and XRFs.

\section{ACKNOWLEDGMENTS}
This work is supported at INAF by funding from 
ASI on grant number I/R/039/04,
at Penn State by NASA contract NASS5-00136 and 
at the University of Leicester by the Particle
Physics and Astronomy Research Council. We
gratefully acknowledge the contribution of dozens 
of members of the XRT team at OAB, PSU, UL, GSFC,
ASDC and our sub-contractors, who helped make 
this instrument possible.

\appendix
\section{\rm XRF~050416A and the Band model}

{\rm O'Brien et al. (2005) have shown that for most GRBs the X-ray afterglow 
light curves smoothly connect with the end of the prompt emission. 
To show this, it is necessary to compute the expected flux in the 
XRT energy range due to the prompt emission.
In Fig.~\ref{lc} we show that the the backward extrapolation 
of the X-ray light curve of XRF~050416A to the time of the trigger 
smoothly joins with the extrapolation of the BAT light curve in the
XRT energy band obtained extending the BAT best fit power law with 
$\Gamma=3.1$ down to soft X-ray energies.
However, the XRF nature of the source implies that the BAT light curve
in Fig.~\ref{lc} represents only an upper limit to X-ray emission
during the prompt phase. 
A BAT light curve extrapolation computed by adopting the Band best fit
model presented by \citet{saka4}, 
or any Band model with $-2 < \alpha_{\rm Band} < -0.5$  
and $E_{\rm p} \sim 16$~keV, predicts a flux significantly below the backwards 
extrapolation of the phase A XRT light curve (more than one order of magnitude).

This would suggest that the steep decline phase is not tied to the prompt
emission, but is, perhaps, due to a further burst peak or a flare (too faint to
be detected by BAT) which occurred during the data gap not covered by XRT.

It is however interesting to note that the spectral index of the phase A XRT
light curve is very soft, and is consistent, within the errors, with that of
the {\em high-energy} slope of the main burst spectrum. 
The phase A light curve could therefore be indeed the prompt emission tail 
(as with most Swift GRBs), provided that the peak energy had shifted red-ward 
of the XRT range by the beginning of the XRT observation. 
Indeed, softening of GRB spectra is a common property of their prompt emission
\citep{ford95,romano06}. \citet{ghir02} showed that in some cases this behavior is due
to the lowering of $E_{\rm p}$. Thanks to Swift, this phenomenon might have 
now been observed over a much wider temporal and spectral span.

\begin{figure}[ht]
\centerline{\includegraphics[width=6cm,angle=-90]{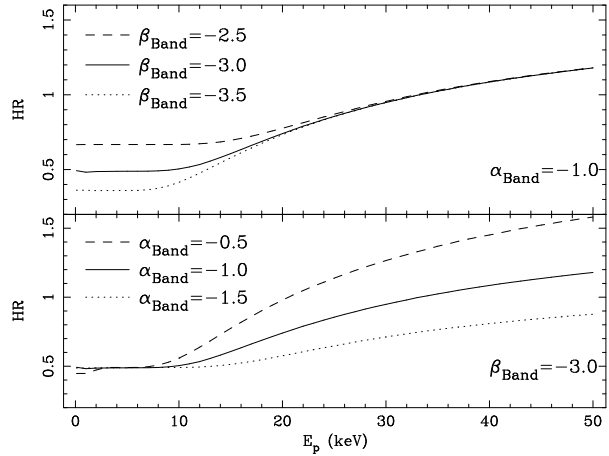}}
\caption{Plots of $(25-50$~keV)/$(15-25$~keV) hardness ratio values
from simulated BAT observations of a source with Band spectral
energy distribution at the same position on the detector array as XRF~050416A.
Top: Hardness ratio as a function of $E_{\rm p}$, for different values of
$\beta_{\rm Band}$ with $\alpha_{\rm Band}$ fixed to $-1$.
Bottom: Hardness ratio as a function of $E_{\rm p}$, for different values of
$\alpha_{\rm Band}$ with $\beta_{\rm Band}$ fixed to $-3$.}
\label{simul}
\end{figure}

The XRF\,050416A BAT spectra showed spectral evolution \citep{saka4}, with the hardness
ratio decreasing during each burst peak. Fig.~\ref{simul} shows the expected
values of the $(25-50$~keV)/$(15-25$~keV) hardness ratio in simulated BAT
observations of a source with a Band spectral energy distribution located at
the same detector position as XRF~050416A. The hardness ratio is plotted as a
function of $E_{\rm p}$ for different choices of the indices $\alpha_{\rm
Band}$ and $\beta_{\rm Band}$, covering their typical ranges of variation
\citep{preece00}.
Simulations of BAT spectra with a Band shape with $\alpha_{\rm Band} = -1$,
$\beta_{\rm Band} = -3$, and $E_{\rm p}$ varying from 0.1 to 50~keV show
that the observed hardness ratio of the XRF\,050416A BAT peaks (which decreased
from $\sim 1$ to $\sim 0.4$) can be reproduced by $E_{\rm p}$ evolving from
$\sim 30$ to $\sim 10$~keV, consistent with the average value $E_{\rm p} \sim
16$~keV \citep{saka4}. A similar behavior could be obtained also allowing an
evolution of $\alpha_{\rm Band}$ 
(from $\alpha_{\rm Band}>-1$ to $\alpha_{\rm Band}<-1$), 
while pure $\beta_{\rm Band}$ evolution (keeping fixed $E_{\rm p} =
16$~keV and $\alpha_{\rm Band} = -1$) cannot easily account for the observed
hardness ratio range.

To summarize, the BAT and XRT data are consistent with the peak energy evolving
from $\sim 30$~keV (during the second GRB peak) down to $\simlt 1$~keV (at the
beginning of the XRT observation). This would roughly correspond to $E_{\rm p}
\propto t^{-1}$. Indeed, fixing the intrinsic absorption to the value obtained
by the fit of the Phase B and phase C spectra, the phase A XRT spectrum may be fit 
by a Band model with $E_{\rm p} \la 1$~keV (although the data do not require this). 
Otherwise, $E_{\rm p}$ could be even lower, close to or below the XRT range 
(so that the $E_{\rm p}$ evolution would be faster).
}

\end{document}